\documentstyle[11pt,asp246,twoside]{article}
\def\edcomment#1{\iffalse\marginpar{\raggedright\sl#1\/}\else\relax\fi}
\reversemarginpar

\begin{document}
\title{Monitoring Variability of the Sky}
\author{Bohdan Paczy\'nski}
\affil{Princeton University, Princeton, NJ 08544-1001, USA}
% \author{Ima Co-Author}
% \affil{The Name of My Institution, The Full Address of My Institution}

\begin{abstract}
Variability in the sky has been known for centuries, even millennia, but
our knowledge of it is very incomplete even at the bright end.  Current
technology makes it possible to built small, robotic optical instruments,
to record images and to process data in real time, and to archive them 
on-line, all at a low cost.  In addition to obtaining complete catalogs 
of all kinds of variable objects, spectacular discoveries can be made, 
like the optical flash associated with GRB 990123 and a planetary transit
in front of HD 209458.  While prototypes of such robotic instruments have
been in operation for several years, it is not possible to purchase a 
complete system at this time.  I expect (hope) that complete systems will
become available `off the shelf' in the near future, as monitoring bright
sky for variability has a great scientific, educational and public outreach
potential.
\end{abstract}

\section{Introduction}

During the last decade several billion dollars have been spent worldwide to
build 6.5-10 meter class telescopes, and there are about 15 of those giants
in operation or under construction.  With ever larger apertures and ever
more sophisticated detectors it is possible to study the universe not only
in ever greater detail but also to make entirely new and very important
discoveries.  However, in this very expensive race to reach the faintest
objects, with the highest angular and spectral resolution and over the
widest spectral range, a broad area of research has been largely neglected: 
the monitoring of optical sky for variability. 

The all sky monitors were known in X-ray and gamma-ray domains for
many decades.  The two examples are: Compton GRO (exists no more)\\
\centerline{ http://cossc.gsfc.nasa.gov/cossc/ } 
\noindent
and Rossi XTE (still in operation)\\
\centerline{ http://heasarc.gsfc.nasa.gov/docs/xte/xte\_1st.html }
\noindent
capable of monitoring gamma-ray and X-ray variability on time scales 
from milliseconds to years over the whole sky.  And there is BACODINE
system with the GCN electronic circulars\\
\centerline{  http://gcn.gsfc.nasa.gov/gcn/gcn\_main.html }
\noindent
which provides
worldwide distribution of `what is new' in the X-ray and gamma-ray sky.  But
there is no such rapid discovery and distribution system in the optical domain.
True, BACODINE/GCN is used by optical and radio observers to report their
follow-up work triggered by X-ray and gamma-ray events.  Also, gravitational
microlensing, supernovae and asteroid searches provide optical alerts for very
limited areas in the sky.  Yet, there is no optical system capable of 
recognizing that something new and unexpected is happening anywhere in the
sky, and to instantly verify the discovery.  The only all sky optical
monitoring is done by amateurs using the naked eye, and therefore it is limited
to 4 mag, or so, with the verification and follow-up possible on a time scale
of hours or days, but not seconds, as it is the case with the BACODINE/GCN.

Professional astronomers do not appreciate how under-explored is the sky 
variability, even at the bright domain.  The ASAS (Pojma\'nski 2000) and
ROTSE (Akerlof et al. 2000) projects demonstrated that by using a 10 cm 
aperture it
is possible to increase the number of known variables brighter than 13 mag
by a factor of 10.  ROTSE (Akerlof et al. 1999) and STARE (Charbonneau et al.
2000) demonstrated that it is possible to make very important discoveries
with such small apertures: the optical flash from a redshift z = 1.6, and the
planetary transit in front of a star.  But note:
these two spectacular discoveries were made in a follow-up mode, with
the target area in the sky, and the target star, selected with
expensive space instruments (BATSE, BeppoSAX) and/or a large optical
telescope.  The existing hardware: small, inexpensive robotic instruments, 
can detect optical flashes and planetary transits, but the software required
to make independent discoveries does not exist, and there appears to be 
little will to develop it, though check McGruder (2001) and STARE:\\
\centerline{ http://www.hao.ucar.edu/public/research/stare/stare.html }

In this presentation I am making a case for small instruments.  There is
a lot of science to be done, but to be efficient, and therefore effective,
the small instruments must be fully robotic.
It is a great pleasure to develop and
operate such instruments: no need to struggle with TACs (time allocation
committees), no need to write observing proposals.  A large team
with all the managerial and funding problems is not necessary, as a full
system can be developed with very modest funds by a competent
individual or a small group, as demonstrated by ASAS (Pojma\'nski 2001):\\
\centerline{ http://archive.princeton.edu/\~~asas/ }
Still, it is not trivial to develop fully robotic hardware, and
robust software is the main bottleneck.  

\section{Today's Systems}

There are several projects which use small robotic instruments to image
almost all sky every clear night, or every few nights.  Almost all of them 
are focused on specific targets, usually searching for optical flashes
associated with gamma-ray bursts, and archiving data with no serious attempt
to analyze it.  The volume is huge, in some cases several terabytes, so
data handling is not easy, and data analysis seems beyond the capability or
interest of large teams involved, like ROTSE, LOTIS, TAROT, STARE.
Links to their Web sites may be found at\\
\centerline{ http://www.astro.princeton.edu/faculty/bp.html }
\centerline{ http://alpha.uni-sw.gwdg.de/\~~hessman/MONET/ }
\noindent

There are many projects taking data, but as far as I know
only one of them monitors everything that varies within its field of
view: the All Sky Automated Survey. Interestingly, ASAS is
a single person undertaking (Pojma\'nski 2000,
2001).  Unfortunately, even ASAS is still processing data off-line.  Another
small group, OGLE (Udalski et al. 1997):\\
\centerline{ http://bulge.princeton.edu/\~~ogle/ }
\noindent
is processing all data almost in real time, but its Early Warning System (EWS)
alerts on microlensing events only, on a time scale of a day or so, and it
monitors less than 0.001 of the sky.  I expect that within a year or so
ASAS's software will be like OGLE's, perhaps even faster, with the alert
time scale of minutes, and near real time verification of anomalous 
photometric and/or
astrometric variation of any type.  Perhaps some other team will reach this
capability ahead of ASAS.  It would be great if this became a common
mode of operation.

Unfortunately, the present small robotic instruments 
are little more than prototypes.  It is not possible to order and
purchase a complete system, or just a complete hardware, for a known price.
The only exception is CONCAM, a very compact camera (cf. Nemiroff et al. 
2000, Pereira et al. 2000, Perez-Ramirez et al. 2000):\\
\centerline{ http://concam.net/ }
capable of imaging 
all sky every few minutes using a CCD detector with a relatively small number
of pixels, controlled by a lap-top computer.  Unfortunately, at this time no
photometric/astrometric pipeline software exists for CONCAM.  In general,
no complete and portable software package is available for any small 
instrument.  This means that most of the data are just archived, and never 
fully processed.
I suppose this is not unusual for a field which is in its early stages of
development, with relatively few people involved, and even fewer people
convinced that it is scientifically useful to have open-minded rather than
narrowly focused observing projects.

\section{Scientific and Educational Goals}

The list of known types of variable objects is long.  It includes eclipsing,
pulsating, and  exploding stars, active galactic nuclei (AGNs), asteroids,
comets, and a large diversity of optical flares or flashes.  Scientific goals 
are very diverse.  Complete catalogs
of variable stars are needed for studies of galactic structure and stellar
evolution.  Calibration of various distance indicators can be done with
the nearest, and therefore apparently the brightest objects. 
AGNs are still poorly understood, and they vary
on all time scales longer than an hour or so.
Comets and asteroids are important for studies of the solar system, 
while `killer asteroids' have a great potential for entertainment, as most
of them are not deadly at all, just spectacular.
Finally, with so many big telescopes in 
operation and under construction it is useful to have a variety of targets
of opportunity detected in real time (cf. Paczy\'nski 1997, 2000, Nemiroff
\& Rafert 1999, and references therein).  

What makes small instruments
scientifically interesting is the very high data rate which can be generated
and processed at low cost, provided suitable software is available.  With
the gradual decrease of detector prices it is possible to have a large number
of pixels.  Computer power is increasing and its cost is falling all the time.
The operating expenses of OGLE hardware translate to over 100,000
photometric measurements per \$1. 
By the time these proceedings are published the cost of 1 terabyte of IDE 
disk will be about \$2,000, making it possible to have huge data sets on-line.
So, it makes sense to use
a `vacuum cleaner' approach and to process all CCD frames and convert 
`pixel data' into `catalog data', which can be analyzed by a much broader
range of astronomers, even amateurs.  The diversity of data types is
small, as all variables are point sources, but the diversity of phenomena
is large, making the database interesting for a variety of scientific as
well as educational projects. 

With an avalanche of data a small team like OGLE or ASAS cannot possibly
analyze it all.  A question comes up: should the data be kept in a closet
for future analysis, or should it be made public domain so other astronomers
can do science with it now?  My view, as well as the view of the OGLE and ASAS
teams, is that the latter solution is preferable; with the data rate increasing
exponentially there will never be time to analyze it all internally.  An 
example of this policy is a recent publication by the OGLE team of almost 
1,000,000 photometric measurements for almost 4,000 Cepheids in the LMC and
SMC (Udalski et at. 1999a,b).  The data posted on the web was analyzed by 
Dr. D. S. Graff of the Ohio State University, who noticed a small but clear
systematic error reaching several hundreds of a magnitude near the
edges of OGLE images.  The error was verified and the
electronic archive was revised on April 1, 2000.  Subsequently, the data
was successfully used to study the geometry of LMC and SMC (Groenewegen 
2000).  While this work was being done elsewhere the OGLE team had time to work on
other projects, and Dr.  A. Udalski had time to work on a new, large CCD
camera for OGLE.

Some focused projects may have very diverse applications.  Let me give
two examples.

The very successful Katzman Automatic Imaging Telescope (Filippenko 2001)
discovers dozens of relatively nearby supernovae every year.
It is not an all sky system, but it monitors several thousand galaxies.
It would be great if the Katzman system could be copied,
and the detection rate of supernovae increased to $ \sim 1,000 $ per year.
Who needs so many events?
One of the most outstanding unsolved problems in modern astrophysics
is a relation between supernovae and gamma-ray bursts (GRBs).
It is likely that GRBs are strongly beamed (cf. Frail
et al.  2001, and references therein).
If the true GRB rate is $ \sim 1,000 $ times higher than the observed rate,
then up to 1\% of all supernovae may generate a gamma-ray
burst which in most cases is not beamed at us.  However, its afterglow may
become detectable as a bipolar radio-supernova remnant several years after the
explosion.  Some supernovae are followed by a strong radio signal which can
be detected with the VLA.  The few sources which are strong enough could be
followed-up with the VLBA with a sub-milli-arcsecond resolution.  The
explosions related to GRBs which are not beamed at us could be recognized
by their bipolar structure and relativistic expansion.  We need as many
nearby supernovae as possible to have a chance to discover those 
hypothetical bipolar radio remnants (Paczy\'nski 2001)

A search for near Earth asteroids (NEAs) is aimed at the discovery of all 
(or at least most) `killer asteroids'.  These are objects which could, upon
impact, cause global catastrophe.  The minimum diameter is estimated to be 
$ \sim 1 $ km.   Hundreds of such asteroids were already discovered by many
projects, like\\
\centerline{ LINEAR, http://www.ll.mit.edu/LINEAR/ }
\centerline{ LONEOS, http://asteroid.lowell.edu/asteroid/loneos/loneos\_disc.html }
\centerline{ NEAT, http://neat.jpl.nasa.gov/ }
\centerline{ SPACEWATCH, http://www.lpl.Arizona.edu/spacewatch/ }
An up to date information may be found at\\
\centerline{ MPC, http://cfa-www.harvard.edu/cfa/ps/mpc.html }
and a recent review was written by Ceplecha et al. (1998).  The searches
discovered also a large number of smaller objects, down to several meter
diameter.  It turns out that about once a month an asteroid 12 meters
across, with a mass of $ \sim 1,000 $ tons collides with Earth and releases
in the upper atmosphere $ \sim 10 $ kilotons of TNT equivalent. 
Ten times more energetic events happen once a year.  These are spectacular
fireballs with strong acoustic effects, but no harm is done at the ground
level.  About once a century a Tunguska-like event releases up to 10 megaton
in an explosion which is locally devastating.  

The searches for near Earth asteroids are done
with 1 meter class telescopes, which implies that a major part of the sky is
covered once every week or so.  This is frequent enough to discover a broad
range of asteroid sizes, and to make statistical estimates of the probability
of impacts, but not frequent enough to recognize those few which are about
to collide with Earth.  Naturally, the huge data archive contains
information about many types of objects with variable brightness, but they
are ignored unless they also change their position.  This is a huge untapped
treasure with the information of general variability of thousands, perhaps
millions of stars and AGNs.  In the next section
I shall discuss modest extensions of the current asteroid searches which
could provide alerts about impending impacts as well as alerts for any unusual
variability in the sky.

Educational opportunities offered by small telescopes are very well described
elsewhere in these proceedings (Hessman 2001).  Unfortunately, neither OGLE nor
ASAS has an active educational program so far.

\section{Prospects for the Future}

With all the current, very diverse activity one goal has not been achieved
so far: we do not know what is happening in the sky in real time, even at the
bright end.  This is a huge gap in astronomical research, which can be
filled only with small, wide angle instruments.  The goal is to monitor all 
sky at the shortest possible time intervals down to whatever magnitude limit 
is technically and financially feasible, to process the data as soon as it is
acquired, to send alerts, and to archive the results in public domain, so
that broad scientific analysis can be accessible to  many users, who
could be called virtual observers.

There are several obvious steps to be made for this idea
to become viable.  First, a complete system with fully automated
hardware and software pipeline and real time alert system should become
operational - none exists at this time.  Next, hardware should be
made easy to duplicate, to allow for a relatively simple expansion
to various sites and various groups.  The availability and the cost of
all the components should be known, and it should be low.  I expect software to
be public domain and free.  Once the systems spread it will be necessary to
find a way to coordinate the data flow, the diversity of alerts and
the ever growing on-line archive.  I think there will be many problems 
which are impossible to predict and we should be open-minded and flexible.
It is very likely that several distinctly different systems will be developed,
with a broad range of costs, data rates, depths and the scope of
surveys.  

Notice that optical variability may have a time scale as short as a 
microsecond, as long as the age of the universe, and anything in between.
It is obviously impossible to cover the whole sky every second to 24 mag.
But it is relatively easy to cover it down to 10 mag every minute, and this
may be a good start, or a good followup on the CONCAM.
The search space is multi dimensional: how large area in the sky is 
monitored, how often, how deep, in which filters, with what accuracy? All 
past and current searches operate in some area of this parameter space, and
there is no way to know where the most spectacular and unexpected discoveries
are to be made.  Consider an example: for over a century enigmatic 
super-flares were observed
on normal main sequence, single, slowly rotating stars of F8 - G8 type
(Schaefer 1989, Schaefer et al. 2000).  However, an instant follow-up
was never done and their nature, or even their reality is not known.
Perhaps they are not actually on the stars, but
on companion planets (Rubenstein 2001)?

A serious discussion is under
way to define science to be done with the LSST (Large Synoptic Survey
Telescope = Dark Matter Telescope, Tyson et al. 2000).  This is a project
to build a telescope with a fast 8.4 meter mirror, a field of view 3 degrees
across, and 1.4 Giga pixels.  If built, it will be able to image all sky
in just a few nights, reaching 24 mag, and saturating at 15 mag.  As powerful
as it will be, LSST will not be likely to discover an optical flash like the
one associated with GRB 990123, as at any given time LSST will image only
$ \sim 0.0001 $ of the sky.  Of course, LSST will
detect a large number of very interesting faint transients.  But note:
it is much easier to follow-up an optical flash which peaks at 10 mag
than one that peaks at 24 mag, yet the bright sky variability on a time scale
of seconds or minutes is not explored at all.

A detection of small asteroids about to collide with Earth should be
possible several hours or even days prior to their impact.  The alerts
would be useful not only scientifically, but they also would be great
for public outreach and entertainment if the time and location of the
next explosive fireball in the upper atmosphere could be predicted.
Such alerts are not possible now, but several near flybys were reported.
A few years ago a graduate student in Tucson, Timothy Spahr, discovered
a 300 meter diameter asteroid 1996 JA1, passing within 450,000 kilometers
of Earth (Spahr 1996).  It reached 11 mag at the closest approach:\\
\centerline{ http://cfa-www.harvard.edu/cfa/ps/mpec/J96/J96K06.html }

A typical relative velocity of an approaching asteroid
is $ \sim 14 $ km/s, which implies 
$ \sim 8 $ hour time to reach Earth from the Moon distance.  A rock with a 
30 meter diameter appears as a $ \sim 15 $ mag object at 400,000 km.
A 12 meter rock would be $ \sim 14 $ mag two hours prior to its impact.
Such objects collide with Earth once per year and once per month, respectively.
Obviously, detecting them is not easy, but it is
not outside the range of current
technology.  If recognized as heading our way the follow-up observations
and the determination of their trajectory would have to be done very quickly,
and presumably automatically, in order to make a prediction of the time and
the location of their impact.  The publicity would be justified, even for
near misses, i.e. near Earth flybys.  A major asteroid or comet on a Tunguska
scale, with a $ \sim 100 $ meter diameter, might be detected several days
prior to its impact, providing enough time to evacuate the `ground zero'.

There is no obvious limit to the expansion of all sky monitoring.  Gradual
reduction of detector and computer costs will make it possible to cover all
sky every night, every
hour, every minute, to ever lower flux limits.  Modest scientific returns
are to be expected even for a project reaching 14 mag every night
(like ROTSE) or 10 mag every minute (a bit better than CONCAM) provided
the data analysis is automated and real time alerts of any unusual variability
are implemented.

\acknowledgments
It is a pleasure to acknowledge the support by NSF grants AST-9819787 and
AST-9820314.

\end{document}